Room-Temperature Electron Spin Transport in a Highly Doped Si Channel

Toshio Suzuki*, Tomoyuki Sasaki[1], Tohru Oikawa[1], Masashi Shiraishi[2],

Yoshishige Suzuki[2], and Kiyoshi Noguchi[1]

AIT, Akita Research Institute of Advanced Technology, Akita 010-1623, Japan

[1]SQ Research Center, TDK Corporation, Saku, Nagano 385-8555, Japan

[2]Graduate School of Engineering Science, Osaka University, Toyonaka, Osaka

560-8531, Japan

Abstract

We report on the first demonstration of generating a spin current and spin transport in a highly doped Si channel at room temperature (RT) using a four-terminal lateral device with a spin injector and a detector consisting of an Fe/MgO tunnel barrier. Spin current was generated using a nonlocal technique, and spin injection signals and Hanle-type spin precession were successfully detected at 300 K, thus proving spin injection with the elimination of spurious signals. The spin diffusion length and its lifetime at RT were estimated to be 0.6 μm and 1.3 ns by the Hanle-type spin precession, respectively.

*E-mail address: toshuzuki@ait.pref.akita.jp

The International Technology Roadmap for Semiconductors lists Si-based spintronics as an emerging research area.[1] The first important issue concerning semiconductor spintronics is the injection of spins from a ferromagnetic conductor into a semiconductor channel, which is noted as a conductivity mismatch problem.[2] The introduction of a sufficiently large spin-dependent interface resistance, such as a tunnel junction and a Schottky barrier, has been theoretically proposed as a possible solution to this problem.[3-6] The second issue is the detection of the injected and accumulated spin. Recently, a four-terminal nonlocal detection technique in a lateral spin valve geometry (NL-MR method) has been demonstrated as an electrical detection method for a spin current.[7] The Hanle-type spin precession measurement in nonlocal detection geometry (4T-Hanle method) was also applied as a powerful tool for proving spin injection and transport.[7,8] Spin current in a Si channel was successfully demonstrated by the 4T-Hanle method, but spin current generation was limited to low temperatures below 150 K.[9-11] On the other hand, a three-terminal Hanle-type spin precession measurement (3T-Hanle method), in which a single magnetic contact is used for both the injection and detection of the spin, was applied to detect the spin accumulation under the injection contact.[12,13] The large spin accumulation at room temperature (RT) using this 3T-Hanle method was an exceptional discovery made by Dash *et al*.[14]

Currently, achieving spin transport at RT in a Si channel is one of the key challenges for Si-spintronics devices. Here, we report on the first demonstration of generating a spin current and spin transport in highly doped Si at RT. Spin transport parameters in the Si channel were investigated using the NL-MR method and the 4T-Hanle method.

The four-terminal lateral spin valve devices used in the study were prepared on a silicon-on-insulator (SOI) substrate with a (100) plane consisting of P-doped Si (100 nm)/silicon oxide (200 nm)/undoped Si wafer. The electron concentration in the P-doped Si layer was determined to be $5 \times 10^{19}$ cm$^{-3}$ by Hall effect measurement. This concentration was the same value as that in our previous studies,[11,15,16] and was also similar to that in the study by Dash *et al.*, where the spin accumulation was firstly achieved in the Si channel (n-type of $1.8 \times 10^{19}$ cm$^{-3}$).[14] An Fe (13 nm)/MgO (0.8 nm) layer was grown on the SOI substrate by molecular beam epitaxy after the native oxide layer on the surface of the SOI substrate was removed using dilute hydrofluoric acid (HF) solution. In this study, we observed $2 \times 1$ and $1 \times 2$ reflection high-energy electron diffraction patterns of the Si surface before the growth of the MgO layer. This was not done in our previous studies.[11,15,16] The values of conductivity in the Si channel were $1.03 \times 10^5$ and $9.52 \times 10^4$ $\Omega^{-1}$m$^{-1}$ measured at 8 and 300 K, respectively. The values of the resistance area product of the tunnel contact were 7.1 and 4.6 K$\Omega\mu$m$^2$ measured at 8 and 300 K, respectively.

Figure 1 shows a schematic cross-sectional view of the lateral device fabricated by electron beam lithography for patterning. The Si channel was fabricated by a mesa-etching technique and two ferromagnetic (FM) electrodes−contact 2 of $0.5 \times 21$ $\mu$m$^2$ and contact 3 of $2 \times 21$ $\mu$m$^2$−were formed by ion milling. Then, a 30-nm-thick SiO$_2$ layer was deposited on the device surfaces, except on the upper faces of the two FM electrodes. Contacts 1 and 4 were holes formed in the SiO$_2$, which were then filled with Al. Finally, the pad electrodes of Au (150 nm)/Cr (50 nm) were fabricated on contacts 1−4 by the lift-off method.

Two measurement methods could be applied to the same measurement sample. One was the NL-MR method in the 4T-geometry, implemented using a standard ac lock-in technique ($f$ = 333 Hz) involving the application of an in-plane magnetic field along the *y*-direction. The ac current was determined using a dc voltage that was required for measuring dc currents for the contact between 1 and 2. The other method was the 4T-Hanle method in the same geometry as that in the NL-MR method involving the application of a magnetic field along the *z*-direction. In both of these measurement methods, a current of 1 mA was injected between contacts 1 and 2, and the output voltage was detected between contacts 3 and 4.

Figure 2 shows the results obtained by the NL-MR method. By subtracting the constant background voltages (195 and 460 μV at 8 and 300 K, respectively) produced by an electric coupling between the electric pads from the raw data, the spin accumulation voltages *ΔV* were obtained and converted into *ΔR* by using an injection current $I_{inject}$ of 1 mA. Steep changes in *ΔR* with clear plateaus were successfully obtained at a temperature of not only 8 K but also 300 K. With the increase in temperature, the field for the transition in *ΔR* shifted slightly to the lower field. Anisotropic magnetoresistance (AMR) hysteresis, as previously reported,[15] was observed in dummy samples. Thus, the changes in *ΔR* can be explained by the magnetization reversal of each FM electrode. Therefore, we can conclude that the changes in *ΔR* resulted from the amount of spin accumulations detected as the spin valve effect. This is evidence of the spin current through the Si channel at 300 K.

The spin diffusion length can be accurately estimated from the gap length dependence of *ΔR*, because the clear plateaus on the *ΔR* curves result from successful anti-parallel magnetization alignment. The inset of Fig. 2 shows that the *ΔR* curve decays exponentially with increasing gap length between contacts 2 and 3. Using the tunnel barrier, *ΔR* decreases as a function of the gap length;[7,17] then, the data is fitted using the following equation:

$$\Delta R = \frac{\Delta V_{NL}}{I_{inject}} = \frac{P^2 \lambda_N}{\sigma S} \exp\left(-\frac{d}{\lambda_N}\right), \quad (1)$$

where $P$ is the spin polarization, $\lambda_N$ is the spin diffusion length, $\sigma$ is the conductivity of the Si, $S$ is the cross section of the Si strip, and $d$ is the gap between the contacts 2 and 3. $\lambda_N$ was estimated to be 2.0 +/- 0.3 and 1.0 +/- 0.3 μm at 8 and 300 K, respectively.

Figure 3 shows the 4T-Hanle signals as a function of the measurement temperature. Each point of the curve is shown by the average of the signals obtained by several measurements. As the temperature increases, the signal becomes smaller and the full width at half maximum intensity widens gradually. However, a reverse of the 4T-Hanle signals is clearly observed by applying the reverse field, even at 300 K. This is strong evidence for the spin current through the Si channel up to 300 K.

The spin diffusion length and other transport parameters were estimated by fitting with the analytical solution of the following function:[7,8]

$$\frac{\Delta V(B_\perp)}{I_{inject}} = \pm \frac{P^2}{e^2 N(E_F) A} \int_0^\infty \varphi(t) \cos(\omega t) \exp\left(\frac{-t}{\tau_{sp}}\right) dt,$$

$$\varphi(t) = \frac{1}{\sqrt{4\pi D t}} \exp\left[-\frac{(x_1 - x_2)^2}{4 D t}\right], \quad (2)$$

where $\omega = g\mu_B B$ is the Larmor frequency, $g$ is the g-factor of electron (using $g = 2$), $\tau_{sp}$ is the spin lifetime, $x_1$ and $x_2$ are integrated over the widths of the injector and detector, and $D$ is the spin diffusion constant. We use the relationship of $\lambda_N = (D\tau_{sp})^{1/2}$, thus $P$, $\lambda_N$, and $\tau_{sp}$ are fitting parameters. The fitting lines are in good agreement with the data, as shown by the example in Fig. 3(b). The spin diffusion lengths at 8 and 300 K were estimated to be 2.00 +/- 0.03 and 0.6 +/- 0.2 μm, respectively. These values are consistent with those obtained by the NL-MR method within an error range, as mentioned above. This indicates that our measurements are highly reliable in the temperature range of 8 to 300 K. Therefore, it is obvious that the spin transport of the band-transferred electrons in the Si channel is demonstrated even at 300 K.

Figure 4 shows the temperature dependence of the obtained transport parameters, together with the 4T-Hanle signals $\Delta V$. With the increase in temperature, $\Delta V$ decreases exponentially while the other transport parameters decrease linearly. As indicated in eq. (2), $\Delta V$ is proportional to the square of the spin polarization, $P^2$. Thus, a reduction in the value of $P$ is believed to lead to the temperature dependence of $\Delta V$. On the other hand, the estimated $P$ of 3.6% at 8 K is twofold higher than that in our previous study,[16] resulting in an enhancement of $\Delta V$. In this study, the surface purification of Si was improved over that in the previous study. Therefore, we infer that $P$ must depend on the MgO/Si interface quality.

The spin lifetimes $\tau_{sp}$ were estimated to be 10.0 +/- 0.2 and 6.3 +/- 0.2 ns at 8 and 100 K, respectively. These values are roughly the same as those obtained in our previous study in the temperature range of 8 to 125 K.[16] This is because both $\tau_{sp}$ and $\lambda_N$ are related to the nature of the Si channel. Furthermore, $\tau_{sp}$ at 300 K was estimated to be 1.3 +/- 0.3 ns in this study. Although the mechanism of the temperature dependence of the spin accumulation phenomena is not yet well elucidated, it is quite reasonable to believe that both the interface- and bulk-originated factors, such as defects and impurities, would influence the spin transport in highly doped Si with the ferromagnetic tunnel junction.

In summary, we demonstrated for the first time the simultaneous spin injection and transport in a highly doped silicon channel at RT, which was confirmed by the nonlocal technique, and spin transport parameters at RT were successfully estimated. The results obtained by two measurement methods—the NL-MR and the 4T-Hanle effect measurement methods—were sufficiently consistent. Although the spin polarization is not yet sufficiently high, higher spin polarization and higher spin voltage can be expected with the realization of a higher-quality MgO/Si interface.

**Acknowledgments**

The authors are grateful to K. Muramoto, Y. Honda, and H. Tomita (Osaka University) for valuable discussions and experimental support, and M. Kubota, Y. Ishida, S. Tsuchida, K. Yanagiuchi, and Y. Tanaka (TDK) for experimental support in sample analysis and fruitful discussions. The authors would also like to thank K. Ouchi (AIT) for his encouragement and discussions. The authors would like to express their thanks to H. Nakanishi, S. Kamata, A. Saito (Akita Prefectural R & D Center), and K. Namba, K. Tagami, and H. Utsunomiya (TDK) for supporting this research.

**References**

1) International Technology Roadmap for Semiconductors, 2004 Edition [http://www.itrs.net/links/2004Update/2004Update.htm].

2) G. Schmidt, D. Ferrand, L. W. Molenkamp, A. T. Filip, and B. J. van Wees: Phys. Rev. B **62** (2000) R4790.

3) E. I. Rashba: Phys. Rev. B **62** (2000) R16267.

4) A. Fert and H. Jaffrès: Phys. Rev. B **64** (2001) 184420.

5) D. L. Smith and R. N. Silver: Phys. Rev. B **64** (2001) 045323.

6) G. E. W. Bauer, Y. Tserkovnyak, A. Brataas, J. Ren, K. Xia, M. Zwierzycki, and P. J. Kelly: Phys. Rev. B **72** (2005) 155304.

7) F. J. Jedema, H. B. Heersche, A. T. Filip, J. J. A. Baselmans, and B. J. van Wees: Nature **416** (2002) 713.

8) X. Lou, C. Adelmann, S. A. Crooker, E. S. Garlid, J. Zhang, K. S. M. Reddy, S. D. Flexner, C. J. Palmstrøm, and P. A. Crowell: Nat. Phys. **3** (2007) 197.

9) O. M. J. van't Erve, A. T. Hanbicki, M. Holub, C. H. Li, C. Awo-Affouda, P. E. Thompson, and B. T. Jonker: Appl. Phys. Lett. **91** (2007) 212109.

10) B. Huang, D. J. Monsma, and I. Appelbaum: Phys. Rev. Lett. **99** (2007) 177209.

11) T. Sasaki, T. Oikawa, T. Suzuki, M. Shiraishi, Y. Suzuki, and K. Noguchi: IEEE Trans. Magn. **46** (2010) 1436.

12) B. Huang, H.-J. Jang, and I. Appelbaum: Appl. Phys. Lett. **93** (2008) 162508.

13) M. Tran, H. Jaffrès, C. Deranlot, J.-M. George, A. Fert, A. Miard, and A. Lemaître: Phys. Rev. Lett. **102** (2009) 036601.

14) S. P. Dash, S. Sharma, R. S. Patel, M. P. de Jong, and R. Jansen: Nature **462** (2009) 491.

15) T. Sasaki, T. Oikawa, T. Suzuki, M. Shiraishi, Y. Suzuki, and K. Tagami: Appl. Phys. Express **2** (2009) 053003.

16) T. Sasaki, T. Oikawa, T. Suzuki, M. Shiraishi, Y. Suzuki, and K. Noguchi: Appl. Phys. Lett. **96** (2010) 122101.

17) S. Takahashi and S. Maekawa: Phys. Rev. B **67** (2003) 052409.

**Figure Captions**

Fig. 1. Schematic cross-sectional view of a Si-based four-terminal lateral device. Magnetic fields were applied along the long axis of the magnetic contacts (*y*-direction) for a nonlocal magnetoresistance measurement and along the *z*-direction for a four-terminal Hanle-type spin precession measurement.

Fig. 2. Nonlocal MR curves at 8 (upper panel) and 300 (lower panel) K, observed in a sample with a gap length *d* of 1.75 μm. The inset shows the gap length dependence of nonlocal magnetoresistance *ΔR*. The red and blue closed circles indicate data at 8 and 300 K, respectively. The dotted lines are the fitting lines.

Fig. 3. Hanle curves observed at different temperatures of the sample with a gap length *d* of 1.75 μm. Hanle curves at (a) 8, 50, 100, 150, 200, 250, and (b) 300 K.

Fig. 4. Temperature dependence, obtained using Hanle-type spin precession, of (a) output voltage *ΔV*, (b) spin polarization *P*, (c) spin lifetime $\tau_{sp}$, and (d) spin diffusion length $\lambda_N$.

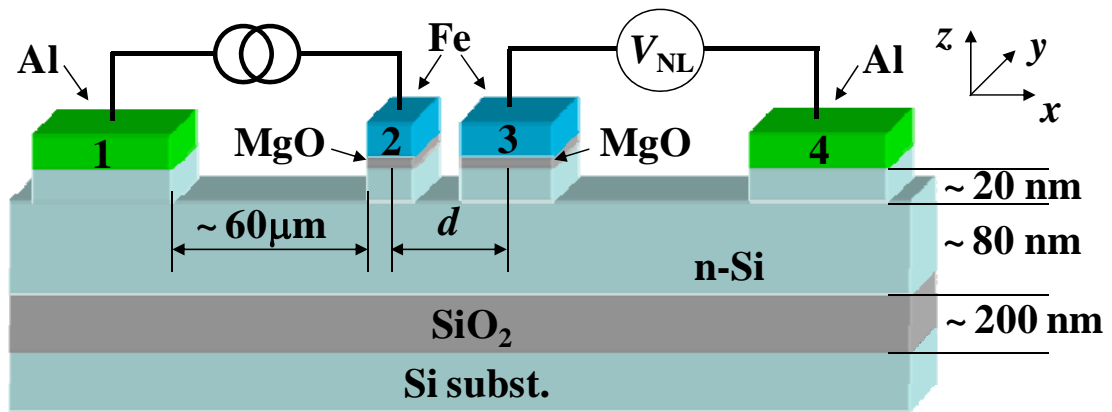

Fig. 1. T. Suzuki *et al*.

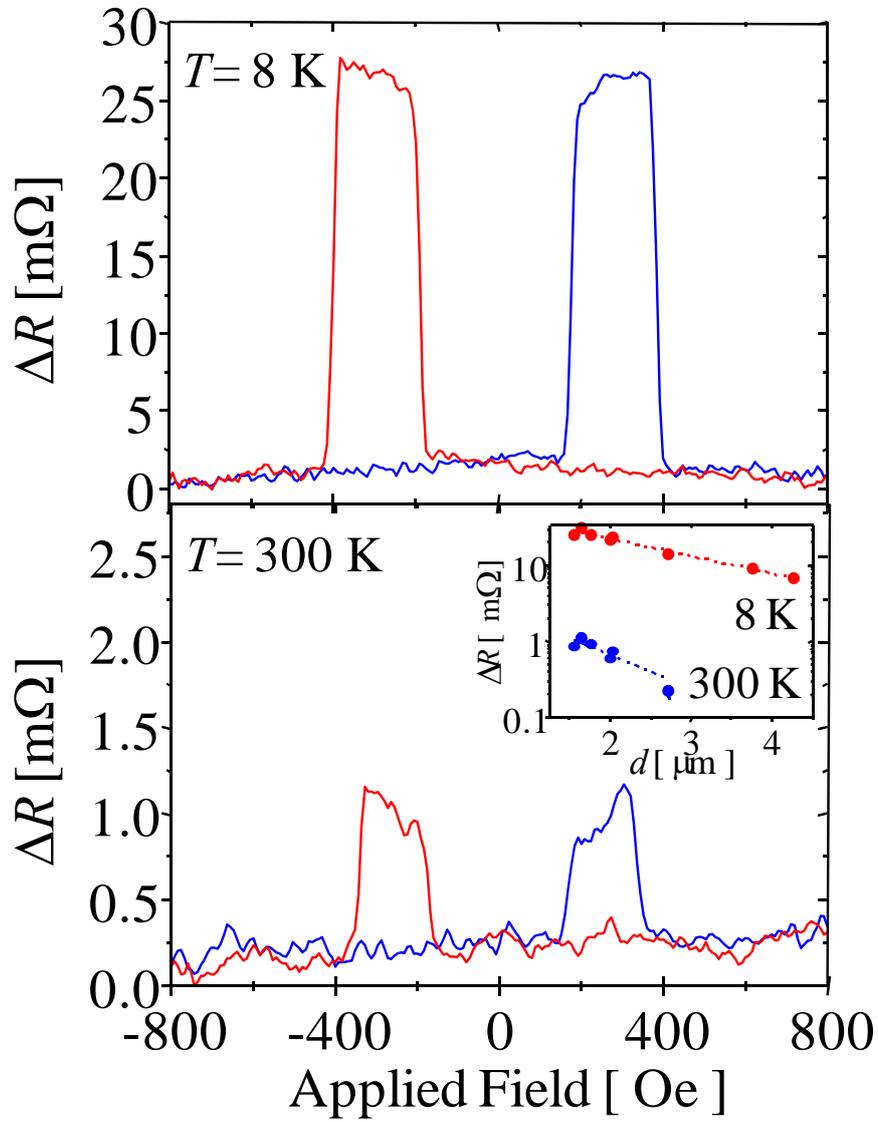

Fig. 2. T. Suzuki *et al*.

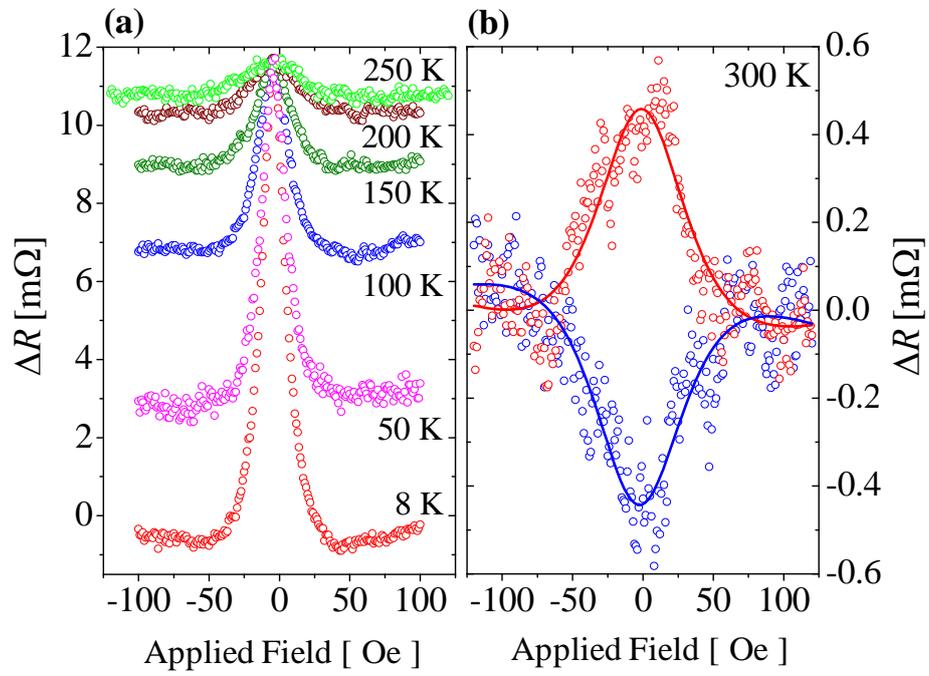

Fig. 3. T. Suzuki *et al.*

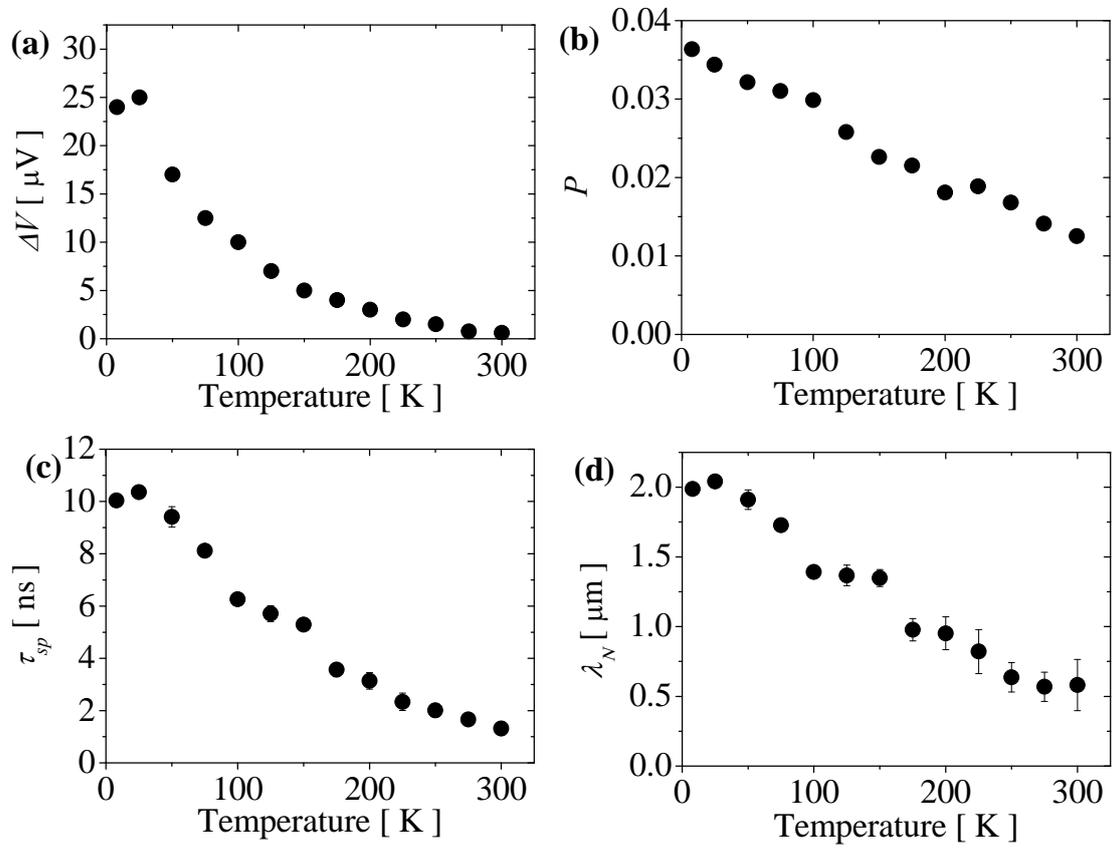

Fig. 4. T. Suzuki *et al*.